\documentclass[prd,aps,superscriptaddress,twocolumn,floatfix,nofootinbib]{revtex4-1}
\pdfoutput=1

\usepackage{amsfonts}
\usepackage{amsmath}
\usepackage{amssymb}
\usepackage{bm}
\usepackage{dcolumn}
\usepackage{graphicx}   
\usepackage[latin1]{inputenc}
\usepackage{latexsym}
\usepackage{rotating}
\usepackage{hyperref}
\usepackage{graphicx}
\usepackage{color}

\newcommand\be{\begin{equation}}
\newcommand\ba{\begin{eqnarray}}
\newcommand\ee{\end{equation}}
\newcommand\ea{\end{eqnarray}}

\begin{document}

\title{Stringy Spacetime Uncertainty Principle and \\
a Modified Trans-Planckian Censorship Criterion}

\author{Robert Brandenberger}
\email{rhb@physics.mcgill.ca}
\affiliation{Department of Physics, McGill University, Montr\'{e}al,
  QC, H3A 2T8, Canada}
 \author{Pei-Ming Ho}
 \email{pmho@phys.ntu.edu.tw}
 \author{Hikaru Kawai}
 \email{hikarukawai@phys.ntu.edu.tw}
 \author{Wei-Hsiang Shao}
 \email{whsshao@gmail.com}
 \affiliation{Department of Physics and Center for Theoretical Physics,
National Taiwan University, Taipei 10617, Taiwan \& Physics Division, National Center for Theoretical Sciences, Taipei 10617, Taiwan.}
 

\begin{abstract}

We study the implications of the stringy space-time uncertainty relation (STUR) for inflationary cosmology. By demanding that no fluctuation modes that exit the Hubble radius are affected by the nonlocality resulting from the STUR, we find an upper bound on the number of e-foldings of inflation. The bound is a factor of $2$ weaker than what results from the Trans-Planckian Censorship Criterion (TCC). By demanding that the inflationary phase is simultaneously consistent with STUR and sufficiently long for inflation to provide a causal explanation of structure on the scale of the current Hubble radius, we find an upper bound on the energy scale of inflation. The bound is less restrictive than what follows from the TCC, but it remains in conflict with canonical single-field inflation models.

\end{abstract}

\maketitle

\section{Introduction} 
\label{sec:intro}

Usual effective field theory approaches based on coupling classical General Relativity to quantum field theory matter are unable to correctly describe the physics deep inside of black holes and the early stages of the universe when the energy density is of the order of the Planck scale. Clearly, a unified quantum theory of space, time, and matter is required, and superstring theory is the most promising approach. It is hence of great interest to study the predictions of string theory for phenomena involving black holes,  and for the very early universe.

A key feature of string theory is the {\em space-time uncertainty relation} (STUR) \cite{STUR}
 \be
 \Delta t \, \Delta x_p \, \geq \, \ell_s^2 \, ,
 \label{STUR-1}
 \ee
where $\Delta t$ and $\Delta x_p$ are the uncertainties in physical time $t$ and physical distance $x_p$, and $\ell_s$ is the string length scale. 
This leads to a
relation between time and physical momentum $k_p$
 \be \label{eq2}
 \Delta t \, (\Delta k_p)^{-1} \, \geq \, \ell_s^2 \, .
 \ee
 
This relation (\ref{eq2}) can also be derived
from the {\em generalized uncertainty principle} (GUP) in the high-energy limit.
The GUP is motivated by string theory \cite{GUP-string},
as well as quantum gravity, in general \cite{GUP-nonstring}.
It is often simply stated as
\be
\label{eq-GUP-space}
\Delta x_p \, \Delta k_p \geq 1 + \ell_s^2 (\Delta k_p)^2,
\ee
which implies a minimal length $\Delta x_p \geq 2 \ell_s$.
In discussions on quantum gravity,
the string length $\ell_s$ is sometimes replaced by the Planck length $\ell_{pl}$ in (\ref{eq-GUP-space}).
For simplicity,
we will not distinguish the Planck scale from the string scale,
that is, $\ell_s = \ell_{pl}$.

The temporal version of (\ref{eq-GUP-space}) is
\be
\label{eq-GUP-time}
\Delta t \, \Delta E \geq 1 + \ell_s^2 (\Delta E)^2,
\ee
where $E$ is the physical energy.
In the high-energy limit, 
the second term on the right-hand side of (\ref{eq-GUP-time}) is much larger than the first term,
leading to 
\be
\label{eq-GUP-HE}
\Delta t \Delta E^{-1} \geq \ell_s^2.
\ee
For a massless particle,
this is equivalent to (\ref{eq2})
due to the on-shell condition $E^2 = k_p^2$.
 
It has recently been shown \cite{Ho1} that this
GUP leads to a shutoff of Hawking radiation after the black hole scrambling time
 \be
 u_{scr} \, = \, 2 r_s {\rm{log}}(r_s m_{pl}) \, ,
 \ee
 where $r_s$ is the Schwarzschild radius, 
$m_{pl} = \ell_{pl}^{-1}$ is the Planck mass,
and $u$ is the Eddington-Finkelstein light-cone time. This, in turn, confirms earlier analyses \cite{Ho2, Ho3} (see also \cite{Ho4, Ho5}), which show on more general grounds that the effective field theory treatment of Hawking radiation will break down on this time scale, a scale which is parametrically shorter than the Page time \cite{Page}.
 
 The question we would like to address in this short article is what the effects of the STUR will be on an accelerating phase of early universe cosmology. Based on the {\it Swampland Program} \cite{swamp1, swamp2} (see \cite{swamprevs} for in-depth reviews), there is mounting evidence that string theory will lead to tight constraints on such an accelerating phase (see e.g. \cite{Vafa-Stein}). Since the STUR is a key result from string theory, it is to be expected that the STUR may also lead to constraints on an accelerating phase of the early universe.
 
 Independent of the Swampland Program,  recently a {\it Trans-Planckian Censorship Conjecture} (TCC) has been formulated \cite{TCC1}.  It states that in no effective field theory coming from quantum gravity,  wavelengths of fluctuation modes that were initially trans-Planckian can ever grow to become larger than the Hubble horizon $H^{-1}(t)$, where $H(t)$ is the Hubble expansion rate.  Applied to a phase of exponential inflation, the TCC implies \cite{TCC2}, as will be reviewed in the next section, an upper bound on the energy scale $\eta$ of inflation of
 \be
\eta \, \leq \, \mathcal{O}(1) \times 10^9 {\rm{GeV}} \, ,
\ee
which is about five orders of magnitude lower than the energy scale of simple inflationary models.  

The TCC is a constraint on models based on usual effective field theory, in which all fields are expanded in comoving Fourier modes, and each Fourier mode is quantized as a harmonic oscillator.  As already pointed out in \cite{Weiss}, there is inevitably a unitarity problem associated with applying this technique in an expanding space.  To avoid the Planck ultraviolet catastrophes, an ultraviolet cutoff must be imposed on the modes, and this cutoff has to be at a fixed physical momentum. To maintain this fixed physical momentum cutoff, there needs to be continuous creation of Fourier modes, a violation of unitarity.  The TCC then follows (see e.g. \cite{RHB-TCC} for a review) by demanding that the non-unitarity does not affect Fourier modes which become super-Hubble, i.e. whose wavelength comes to exceed the Hubble radius $H^{-1}(t)$.

The uncertainty relation (\ref{eq2}) stemming from the STUR can be viewed as a proposal for the applicability of an effective field theory description of a model consistent with string theory. It implies that if we are studying processes that occur on a time scale $\Delta t$, then the usual effective field theory description is applicable, provided we consider modes with a wavenumber consistent with (\ref{eq2}).  Here, we wish to explore what the implications of this constraint are for an inflationary phase of the early universe. We find a modified and somewhat weaker version of the TCC.
 
Our work is complementary to what of \cite{Matteo} in which the implications of the cutoff frequency imposed in \cite{Ho3} on the near horizon modes were studied for the static patch description of de Sitter. It was found that on twice the TCC time scale, the cutoff prescription suggested by the STUR  leads to a shut-off of de Sitter radiation, in analogy to how black hole Hawking radiation shuts off in the analysis of \cite{Ho3}.  The analysis in \cite{Matteo} was done by computing the Bogoliubov mode mixing coefficients which determine late-time radiation, assuming the cutoff frequency prescription of \cite{Ho3}. Here, instead, we work in the usual cosmological space-time coordinates and base our analysis on a reformulation of the STUR, which is appropriate for the study of cosmological fluctuation modes.
 
Our work is also related to that of \cite{Ho-Br}, where the space-time uncertainty relation was applied to the study of cosmological fluctuations in an expanding universe.  A mode with wavenumber $k$ was taken to emerge in its ground state at the time $t_i(k)$ when, for the corresponding mode, the STUR is satisfied, thus leading to changes of the spectrum of fluctuations compared to what is obtained in the usual effective field theory analysis. Here, we will, in fact, argue that the demand of being insensitive to the stringy uncertainty scale leads to an upper bound on the duration of a phase of accelerated expansion.
 
We assume a spatially flat space-time with metric
\be
ds^2 \, = \, dt^2 - a^2(t) d{\bf{x}}^2 \, ,
\ee
where $t$ is physical time, ${\bf{x}}$ are the spatial comoving coordinates, and $a(t)$ is the scale factor, and we work with natural units in which $c = \hbar = k_B = 1$.

\section{Analysis}
\label{section2}

\subsection{Usual TCC Analysis}

We first review the derivation \cite{TCC2} of the upper bound on the energy scale of inflation from the TCC \cite{TCC1}. The TCC implies that the comoving scale whose physical length at the time $t_i$, the beginning of the period of inflation, equals the Planck length cannot evolve to become larger than the Hubble horizon scale $H^{-1}(t_R)$ at $t_R$, the end of the period of inflation.\footnote{If this condition is satisfied, the scale will always remain smaller than the Hubble horizon after inflation. Note that we are here assuming exponential expansion with $H(t)$ constant between $t_i$ and $t_R$. In the case of power-law accelerated expansion, the bound on the energy scale is slightly weaker \cite{weaker}. On the other hand, the bound is in fact stronger if we take pre-inflationary evolution into account \cite{stronger}.} In terms of equations, this condition reads
\be \label{UB}
\frac{a(t_R)}{a(t_i)} \ell_{pl} \, < \, H^{-1}(t_R) \, \equiv \, H^{-1} \, .
\ee
On the other hand, for inflation to be able to provide a causal mechanism for structure formation \cite{Mukh}, the current Hubble horizon scale $t_0$ (where $t_0$ denotes the present time) must originate inside of the Hubble radius at the beginning of inflation. This leads to a lower bound on the duration of inflation, the condition being given by
\be \label{LB}
\frac{a(t_i)}{a(t_0)} t_0 \, < \, H^{-1} \, .
\ee
The geometry of these bounds can be seen in the space-time sketch of Figure 1. Here, the vertical axis is time, and the horizontal axis indicates physical length. The inflationary phase lasts from $t = t_i$ to $t = t_R$.  We assume a fast transition to the expanding radiation phase of standard cosmology.  The red curve labeled $\lambda_1(t)$ indicates the wavelength of the mode which starts with Planck length (indicated by a black vertical line) at the beginning of inflation. In the diagram, the upper bound (\ref{UB}) on the duration of the inflationary phase is shown to be marginally satisfied in that it hits the Hubble horizon (the brown vertical line) at the end of inflation. The blue curve labeled $\lambda_2(t)$ shows the wavelength of the mode which enters the Hubble horizon at the present time $t_0$. If inflation is to provide the origin of structure on the current Hubble horizon scale, the mode has to originate inside of the Hubble horizon at the beginning of inflation. This condition is also shown to be marginally satisfied in the figure. Whether the two conditions can be simultaneously met depends on the ratio of the Hubble horizon length $H^{-1}$ to the Planck length. The lower the energy scale of inflation, the larger this ratio, and the easier it is to simultaneously obey the conditions (\ref{UB}) and (\ref{LB}).

\begin{center}
\begin{figure}[!htb]
\includegraphics[width=6.4cm]{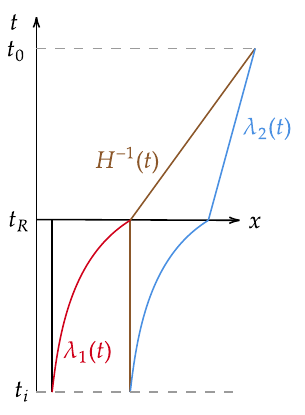}
\caption{Space-time sketch of an inflationary universe. The vertical axis is time, with the interval $t_i < t < t_R$ being the period of inflation. The horizontal axis indicates physical distance.   The curve $\lambda_1(t)$ indicates a fixed comoving scale whose physical wavelength equals the Planck length at $t_i$, which must remain inside the Hubble radius at all times and, in particular, at the end of the period of inflation. The curve $\lambda_2(t)$ is the physical wavelength corresponding to the current Hubble radius which must emerge from inside of the Hubble radius at the beginning of inflation.  In this sketch, the upper and lower bounds on the duration of inflation are shown to be marginally satisfied. This implies that the Hubble radius during inflation must be several orders of magnitude larger than what is usually assumed in single-field inflation models. (This figure is reproduced from \cite{Review}.)}
  \label{fig7}
\end{figure}
\end{center}

For exponential inflation, the ratio of scale factors in (\ref{UB}) can be expressed in terms of the number $N$ of e-foldings of inflation,  and (\ref{UB}) becomes
\be \label{UB2}
N \, < \, {\rm{ln}}\hskip-0.1em\left(\frac{m_{pl}}{H}\right) \, .
\ee
On the other hand,  the condition (\ref{LB}) can be written in the form
\be \label{LB2}
N \, > \, {\rm{ln}}\hskip-0.1em\left(\frac{T_0}{T_R} t_0 H\right) \, ,
\ee
where $T_0$ is the present temperature of the cosmic microwave background, $T_R$ is the temperature at the beginning of the radiation phase of standard Big Bang cosmology expansion, and we have assumed entropy conservation after $T_R$ to replace the radio of scale factors with the ratio of inverse temperatures.

To see under which conditions the two bounds (\ref{UB2}) and (\ref{LB2}) can be consistent, we use the Friedmann equation to express $t_0$ in terms of $T_0$ noting that the energy density today is 
\be \label{present}
\rho_0 \, \simeq \, \frac{\pi^2}{30} g_* T_0^4 \frac{T_{eq}}{T_0} \, ,
\ee
where $T_{eq}$ is the temperature at the time $t_{eq}$ of equal matter and radiation %
and $g_*$ is the number of spin degrees of freedom in the thermal bath at $T_{eq}$ (we are using the usual notation of standard cosmology). The last factor in (\ref{present}) takes into account the fact that the energy density after $t_{eq}$ scales as $a^{-3}$ and not as $a^{-4}$ as the energy density in radiation. 
Using the Friedmann equation to relate $t_0$ to $\rho_0$ (and neglecting the small change in the result which stems from the contribution of Dark Energy to the expansion rate) yields
\be
t_0^2 \, = \, \frac{5}{\pi^3 g_*} m_{pl}^2 \frac{T_0}{T_{eq}} T_0^{-4} \, .
\ee
Hence, (\ref{LB2}) becomes
\be \label{LB3}
N \, > \, {\rm{ln}}\hskip-0.1em\left(\frac{m_{pl}}{H}\right) + {\rm{ln}}\hskip-0.2em\left[ \alpha \left( \frac{T_0}{T_{eq}}\right)^{1/2} \left( \frac{H}{m_{pl}}\right)^{3/2} \frac{m_{pl}}{T_0} \right] \, ,
\ee
where we have made use of the assumption of instantaneous (on the Hubble time scale) reheating
and applied the Friedmann equation to express $T_R$ in terms of $H$:
\be
T_R \, = \, \left( \frac{45}{4\pi^3 g(t_R)} \right)^{1/4} H^{1/2} m_{pl}^{1/2} \, ,
\ee
and $g(t_R)$ is the number of spin degrees of freedom in the radiation bath after inflation. The constant $\alpha$ in (\ref{LB3}) is
\be
\alpha \, = \, \left(  \frac{20 g(t_R)}{9 \pi^3 g_*^2}  \right)^{1/4} \, .
\ee
Hence, the condition for compatibility of the upper and lower bounds on the duration of inflation is that the argument inside the second logarithm in (\ref{LB3}) is smaller than $1$. This yields the criterion
\be \label{bound1}
\frac{H}{m_{pl}} \, < \, \left( \frac{T_0}{m_{pl}} \right)^{2/3} \left( \frac{T_{eq}}{T_0}\right)^{1/3} \alpha^{-2/3}
\ee
on the value of $H$ during inflation. Equivalently, this is an upper bound of
\be
\frac{\eta}{m_{pl}} \, < \, 
\left( \frac{T_0}{m_{pl}}\right)^{1/3} \left(\frac{T_{eq}}{T_0} \right)^{1/6} 
\alpha^{-1/3} \left( \frac{3}{8\pi}\right)^{1/4}
\ee
on the energy scale $\eta$ during inflation. Inserting the values of $m_{pl}$, $T_{eq}$, and $T_0$ we find the constraint
\be
\eta \, < \, {\mathcal{O}}(1) \times 10^9 {\rm{GeV}} \, ,
\ee
which is more than 5 orders of magnitude lower than the value for simple scalar field inflation models \cite{TCC2}.

\subsection{Constraints from the STUR}

The TCC criterion (\ref{UB}) comes from demanding that no trans-Planckian modes ever exit the Hubble horizon. The justification is that for modes that violate the TCC criterion, it is clear that the usual effective field theory analysis will break down. The STUR can be viewed as an attempt to extend the applicability of usual field theory methods to trans-Planckian scales.  Specifically, the trans-Planckian effect is expected to be dominant whenever the condition (\ref{eq2}) is violated.  We here propose an application of (\ref{eq2}) to cosmology.  We will take $\Delta t$ to be 
\be 
\Delta t \, = \, H^{-1} 
\ee
since this is the time scale on which time changes occur in an expanding universe.  For cosmological fluctuation modes with wavenumber $k$, the value of $\Delta x_p$ is set by the physical wavelength. Hence, the cosmological version of the stringy STUR (\ref{STUR-1}) reads
\be \label{new}
H^{-1}(t)  \left( \frac{k}{a(t)} \right)^{-1} \, \geq \, \ell_s^2 \, .
\ee
From the point of view of this version of the STUR, the validity of the usual effective field theory analysis of inflation is that this modified uncertainty relation be obeyed at the initial time $t_i$ of inflation for all modes that eventually leave the Hubble horizon, i.e.,
\be \label{crit}
H^{-1} \left( \frac{k}{a(t_i)} \right)^{-1} \, \geq \, \ell_s^2 \, .
\ee
In particular, this condition must be obeyed for the mode which exits the Hubble horizon at the end of inflation, i.e. for which
\be \label{Hubblemode}
k \, = \, H(t_R) a(t_R) \, .
\ee
Inserting (\ref{Hubblemode}) into (\ref{crit}) leads to the condition
\be \label{crit2}
\frac{a(t_R)}{a(t_i)} \, \leq \, \left( \frac{m_{pl}}{H} \right)^2 \, ,
\ee
recalling that we have set $m_{pl} = \ell_s^{-1}$ for simplicity.
It leads to an upper bound on the e-folding number $N$ which is weaker by a factor of $2$ than the TCC bound of \cite{TCC2}, namely \footnote{Note that it is the number of e-foldings which is larger by a factor of $2$, while in the analysis of \cite{Matteo} it was the time interval itself which was larger by a factor of $2$ than the TCC time scale of \cite{TCC2}.}
\be \label{crit3}
N \, < \, 2 \, {\rm{ln}}\hskip-0.1em\left(\frac{m_{pl}}{H}\right) \, .
\ee

The lower bound on the duration of inflation remains unchanged. It can be written as
\be \label{LB4}
N \, > \, {\rm{ln}}\hskip-0.1em\left(\frac{m_{pl}^2}{H^2}\right) + {\rm{ln}}\hskip-0.2em\left[ \alpha \left( \frac{T_0}{T_{eq}}\right)^{1/2} \left( \frac{H}{m_{pl}}\right)^{5/2} \frac{m_{pl}}{T_0} \right] \, ,
\ee
and the compatibility condition between (\ref{crit3}) and (\ref{LB4}) now is that the argument in the second term on the right hand side of (\ref{LB4}) is smaller than 1, i.e.
\be
\frac{H}{m_{pl}} \, < \, \left( \frac{T_0}{m_{pl}} \right)^{2/5} \left( \frac{T_{eq}}{T_0}\right)^{1/5} \alpha^{-2/5} \, ,
\ee
which is a much weaker condition than the TCC relation (\ref{bound1}) and allows for larger values of $H$. As in the previous subsection, we can express the above as a bound on the energy scale of inflation, and we obtain
\be
\eta \, < \, {\cal{O}}(1) \, 10^{13} {\rm{GeV}} \, ,
\ee
a bound which is still in conflict with simple scalar field toy models of inflation.

\section{Discussion}  

In this letter we have studied the consequences of the stringy space-time uncertainty relation (STUR) on inflationary cosmology. By demanding that a proposed cosmological version of the STUR is obeyed at the beginning of the period of inflation for cosmological fluctuation modes which exit the Hubble horizon at the end of inflation, we have derived an upper bound on the duration of inflation. The resulting maximal number of e-foldings of inflation is twice what is obtained using the TCC criterion of \cite{TCC1, TCC2}.  We call this the {\it modified TCC time scale}.

Note that the time scale we have obtained is parametrically smaller than the ``quantum break time'' 
\be
T_{break} \, \sim \, \frac{m_{pl}^2}{H^2} \frac{1}{H}
\ee
of de Sitter space \cite{Dvali} when de Sitter space becomes unstable (this is the same time scale when the back-reaction of cosmological perturbations on the background becomes dominant \cite{RHB-BR}).  The hierarchy between the TCC and the quantum break time scales is analogous to the hierarchy between the ``scrambling time'' and the ``Page time'' for black holes.

It would be interesting to study the reason for the difference by a factor of 2 in the allowed e-folding number of inflation between the result of our current analysis and what is obtained using the TCC,  The reason why our modified time scale is longer is that we are no longer requiring that those modes which cross the Hubble horizon before the end of inflation are cis-Planckian for all times,  but only that the effects of the space-time uncertainty relation on these fluctuation modes remains small at all times.

String theory is much richer than an effective field theory supplemented by the STUR. Hence, it is likely that the full string theory leads to much stronger constraints on a phase of accelerated expansion than what our analysis has yielded. In fact, the de Sitter criterion \cite{swamp2} appears to completely prohibit standard single field slow-roll inflation, assuming that this phase can be described by effective field theory techniques (see also \cite{no-go} for arguments which rule out de Sitter in the corner of string theory when effective field theory is a good description).  There are attempts to construct a de Sitter phase as a coherent state in M-theory \cite{Keshav}. Such a phase appears to have a duration bounded from above. It would be very interesting to check if this bound obtained in \cite{Keshav} agrees with the TCC bound or the bound we have established.

\begin{acknowledgements}

We thank Samuel Laliberte, Ronny Chau, Henry Liao, Nobuyoshi Ohta, and Cheng-Tsung Wang for valuable discussions.  The research of R.B. at McGill is supported in part by
funds from NSERC and from the Canada Research Chair program.
P.M.H. and W.H.S. are supported in part 
by the Ministry of Science and Technology, R.O.C.
(MOST 110-2112-M-002-016-MY3),
and by the National Taiwan University. 
H.K. thanks Prof. Shin-Nan Yang and his family
for their kind support through the Chin-Yu chair professorship.
H.K. is partially supported by the Japan Society of Promotion of Science (JSPS),
Grants-in-Aid for Scientific Research (KAKENHI)
Grants No.\ 20K03970 and 18H03708,
by the Ministry of Science and Technology, R.O.C. (MOST 111-2811-M-002-016),
and by National Taiwan University.  R.B. wishes to thank Professor Ho for the invitation to visit NTU and the members of his group for their hospitality.

\end{acknowledgements}


\end{document}